\documentclass[a4paper,11pt]{article}
\pdfoutput=1
\usepackage{pos}

\title{Evaluating scattering amplitudes with py\textsc{SecDec}~1.6}

\author*[a]{Vitaly Magerya}
\affiliation[a]{Institute for Theoretical Physics, Karlsruhe Institute of Technology,\\
Wolfgang-Gaede-Str. 1, Geb. 30.23, 76131 Karlsruhe, Germany}
\emailAdd{vitalii.maheria@kit.edu}

\abstract{
    py\textsc{SecDec} is a computer tool to evaluate Feynman
    integrals and their weighted sums (amplitudes) using the
    method of sector decomposition and numerical integration. The
    new release of py\textsc{SecDec} version 1.6 comes with a
    significant performance boost (3x--9x in common scenarios),
    and new features to make the evaluation and asymptotic
    expansion of amplitudes and integrals easier and faster. In
    this article we briefly review these features.
}

\FullConference{%
    16th International Symposium on Radiative Corrections: \\
    Applications of Quantum Field Theory to Phenomenology (RADCOR2023)\\
    28th May -- 2nd June, 2023\\
    Crieff, Scotland, UK\\}

\usepackage{amsmath}
\usepackage{booktabs}
\usepackage{multirow}
\usepackage{prettyref}
\usepackage{xcolor}
\usepackage[absolute]{textpos}

\input{all.tikzdefs}

\tikzstyle{blank}=[fill=white, shape=circle, draw=white, inner sep=0.8pt]
\tikzstyle{dot}=[fill=black, shape=circle, draw=black, inner sep=0.8pt]
\tikzstyle{fat}=[fill=white, shape=circle, draw={rgb,255: red,176; green,36; blue,39}, dashed, line width=1pt, inner sep=0.8pt]

\tikzstyle{edge}=[-, draw=PropagatorColor, line width=1.2pt, line cap=rect, preaction={{draw=white,line width=2.5pt}}]
\tikzstyle{incoming edge}=[line width=1pt, line cap=rect, draw={rgb,255: red,102; green,102; blue,102}, {|-}]
\tikzstyle{outgoing edge}=[line width=1pt, line cap=rect, draw={rgb,255: red,102; green,102; blue,102}, ->]
\tikzstyle{external edge}=[line width=1pt, line cap=rect, draw={rgb,255: red,102; green,102; blue,102}, -]
\tikzstyle{top}=[-, draw=TopPropagatorColor, line width=1.9pt, line cap=rect, preaction={{draw=white,line width=2.5pt}}]
\tikzstyle{edge dot1}=[-, postaction=decorate, decoration={markings,mark=at position .50 with {\node[style=dot]{};}}]
\tikzstyle{edge dot2}=[-, postaction=decorate, decoration={markings,mark=between positions 0.33 and 0.67 step 0.33 with {\node[style=dot]{};}}]
\tikzstyle{edge dot3}=[-, postaction=decorate, decoration={markings,mark=between positions 0.25 and 0.76 step 0.25 with {\node[style=dot]{};}}]
\tikzstyle{edge dot4}=[-, postaction=decorate, decoration={markings,mark=between positions 0.20 and 0.81 step 0.20 with {\node[style=dot]{};}}]
\tikzstyle{dot1}=[-, postaction=decorate, decoration={markings,mark=at position .50 with {\node[style=dot]{};}}]
\tikzstyle{dot2}=[-, postaction=decorate, decoration={markings,mark=between positions 0.33 and 0.67 step 0.33 with {\node[style=dot]{};}}]
\tikzstyle{dot3}=[-, postaction=decorate, decoration={markings,mark=between positions 0.25 and 0.76 step 0.25 with {\node[style=dot]{};}}]
\tikzstyle{dot4}=[-, postaction=decorate, decoration={markings,mark=between positions 0.20 and 0.81 step 0.20 with {\node[style=dot]{};}}]
\tikzstyle{incoming}=[line width=1pt, line cap=rect, draw=LegColor, {|-}]
\tikzstyle{outgoing}=[line width=1pt, line cap=rect, draw=LegColor, ->]
\tikzstyle{outgoing top}=[line width=1pt, line cap=rect, draw=TopLegColor, ->]
\tikzstyle{outgoing higgs}=[line width=1pt, line cap=rect, draw=HiggsLegColor, ->]
\tikzstyle{massive edge}=[-, draw={rgb,255: red,133; green,119; blue,181}, line width=2.0pt, preaction={{draw=white,line width=2.5pt}}, line cap=rect]
\tikzstyle{cut edge}=[-, draw={rgb,255: red,64; green,64; blue,64}, line width=0.5pt, densely dashed, line cap=rect]
\tikzstyle{xcut edge}=[-, draw={rgb,255: red,64; green,64; blue,64}, line width=0.5pt, densely dashed, line cap=rect, postaction={decorate, decoration={markings,mark=at position .50 with {\node[cross out,solid,draw=white,line width=2pt,inner sep=1.4pt,transform shape] {};}}}, postaction={decorate, decoration={markings,mark=at position .50 with {\node[cross out,solid,draw={rgb,255: red,176; green,36; blue,39},line width=1pt,inner sep=1.8pt,transform shape] {};}}}]
\tikzstyle{xcut edge 1/3}=[-, draw={rgb,255: red,64; green,64; blue,64}, line width=0.5pt, densely dashed, line cap=rect, postaction={decorate, decoration={markings,mark=at position .33 with {\node[cross out,solid,draw=white,line width=2pt,inner sep=1.4pt,transform shape] {};}}}, postaction={decorate, decoration={markings,mark=at position .33 with {\node[cross out,solid,draw={rgb,255: red,176; green,36; blue,39},line width=1pt,inner sep=1.8pt,transform shape] {};}}}]
\tikzstyle{cut}=[-, draw={rgb,255: red,61; green,171; blue,83}, line width=0.7pt, dotted, line cap=rect]
\tikzstyle{photon}=[-, draw={rgb,255: red,176; green,36; blue,39}, line width=1pt, preaction={{draw=white,line width=2pt}}, line cap=rect, decorate, decoration=snake]
\tikzstyle{gluon}=[-, draw={rgb,255: red,176; green,36; blue,39}, line width=1pt, preaction={{draw=white,line width=2pt}}, line cap=rect, decorate, decoration={coil,aspect=1.4,segment length=2.5mm}]
\tikzstyle{gluoncoil}=[-, decorate, decoration={coil,aspect=1.4,segment length=2.5mm}]
\tikzstyle{fermion}=[-, draw={rgb,255: red,176; green,36; blue,39}, line width=1pt, preaction={{draw=white,line width=2pt}}, line cap=rect, postaction=decorate, decoration={markings,mark=at position .60 with {\arrow{stealth[round]}}}]
\tikzstyle{ghost}=[-, style=fermion, line width=1pt, line cap=round, dash pattern={on 0pt off 3\pgflinewidth}]
\tikzstyle{scalar}=[-, line width=1pt, densely dashed, draw={rgb,255: red,102; green,102; blue,102}]
\tikzstyle{fermionarrow}=[-,postaction=decorate, decoration={markings,mark=at position .60 with {\arrow{stealth[round]}}}]
\tikzstyle{brace}=[-,draw={rgb,255: red,61; green,171; blue,83}, line width=1pt, decorate, decoration={brace,amplitude=5pt}]

\tikzstyle{edge}=[-, draw=EdgeColor, line width=1.2pt, line cap=rect, style=whitebg]
\tikzstyle{incoming edge}=[|-, style=edge, draw=PaleEdgeColor, style=arrowin]
\tikzstyle{outgoing edge}=[->, style=edge, draw=PaleEdgeColor, style=arrow]

\tikzstyle{massive edge}=[-, draw=MassiveEdgeColor, line width=2pt, style=whitebg, line cap=rect]
\tikzstyle{incoming massive edge}=[|-, style=massive edge, draw=PaleMassiveEdgeColor, style=arrowin]
\tikzstyle{outgoing massive edge}=[->, style=massive edge, draw=PaleMassiveEdgeColor, style=arrow]

\tikzstyle{massive2 edge}=[-, draw=Massive2EdgeColor, line width=2pt, style=whitebg, line cap=rect]
\tikzstyle{incoming massive2 edge}=[|-, style=massive2 edge, draw=PaleMassive2EdgeColor, style=arrowin]
\tikzstyle{outgoing massive2 edge}=[->, style=massive2 edge, draw=PaleMassive2EdgeColor, style=arrow]

\tikzstyle{massive3 edge}=[-, draw=Massive3EdgeColor, line width=2pt, style=whitebg, line cap=rect]
\tikzstyle{incoming massive3 edge}=[|-, style=massive3 edge, draw=PaleMassive3EdgeColor, style=arrowin]
\tikzstyle{outgoing massive3 edge}=[->, style=massive3 edge, draw=PaleMassive3EdgeColor, style=arrow]

\tikzstyle{scalar}=[-, draw=ScalarColor, line width=1pt, densely dashed]
\tikzstyle{incoming scalar}=[-, style=scalar, draw=PaleScalarColor]
\tikzstyle{outgoing scalar}=[-, style=scalar, draw=PaleScalarColor]

\tikzstyle{massive scalar}=[-, draw=ScalarColor, line width=2pt, densely dashed]
\tikzstyle{incoming massive scalar}=[-, style=massive scalar, draw=PaleScalarColor]
\tikzstyle{outgoing massive scalar}=[->, style=massive scalar, draw=PaleScalarColor, style=arrow]

\definecolor{EmeraldGreen}{HTML}{1ea78d}
\definecolor{EnglishRed}{HTML}{b02427}
\hypersetup{
    pdftitle={Evaluating scattering amplitudes with py\textsc{SecDec}~1.6},
    pdfauthor={Vitaly Magerya}
}

\definecolor{ZetaLightBrown}{HTML}{bf8040}

\let\origref\ref
\newrefformat{sec}{\hyperref[#1]{Section~\origref*{#1}}}
\newrefformat{tab}{\hyperref[#1]{Table~\origref*{#1}}}
\newrefformat{fig}{\hyperref[#1]{Figure~\origref*{#1}}}
\newrefformat{eq}{eq.\,\hyperref[#1]{(\origref*{#1})}}
\newcommand{\subfigref}[2]{\hyperref[#1]{Figure~\origref*{#1}{#2}}}
\AtBeginDocument{\renewcommand*{\ref}[1]{\prettyref{#1}}}

\usepackage[absolute]{textpos}
\AtBeginDocument{\begin{textblock*}{15cm}[1,0](20.6cm,8.8cm)\raggedleft\small\texttt{ KA-TP-24-2023}\end{textblock*}}

\begin{document}
\tableofcontents
\maketitle

\section{Introduction}

\noindent Matching the increasingly precise experimental
measurements at the Large Hadron Collider and other colliders on
the theoretical side requires the calculation of higher-order
corrections to scattering amplitudes. One of the key elements
of that is the calculation of multi-loop Feynman integrals;
this is a challenging task, and analytic expressions for many
phenomenologically relevant classes of integrals at two loops
and beyond are not available---instead numeric and semi-analytic
methods are used.

One of the well established methods of numerical evaluation of Feynman
integrals is sector decomposition~\cite{Heinrich:2008si,Binoth:2000ps}.
This method has been continuously refined over the years, and so
have been its implementations, the most prominent of which are
py\textsc{SecDec}~\cite{HJKMOS23,Heinrich:2021dbf,Borowka:2018goh,Borowka:2017idc}
(together with its predecessor \textsc{SecDec}~\cite{Borowka:2015mxa,Borowka:2012yc})
and \textsc{Fiesta}~\cite{Smirnov:2021rhf}.


Recently py\textsc{SecDec} version~1.6 has been
released~\cite{HJKMOS23}.\footnote{One can find py\textsc{SecDec}
documentation at \url{https://secdec.readthedocs.io}, and
the source code repository together with the examples at
\url{https://github.com/gudrunhe/secdec}.} In this article we
shall walk through the new features of this release, their
motivations and benefits, demonstrating the expected performance
and capabilities of py\textsc{SecDec}. These new features include:

\begin{itemize}
\item
    a new evaluator codenamed \textsc{Disteval}, achieving a 3x--9x
    speedup across a wide variety of amplitudes and integrals,
    and providing the possibility of integration distributed
    over multiple computers (\ref{sec:disteval});
\item
    a new probabilistic method of constructing Quasi Monte Carlo
    (QMC) lattices used in integration, called \textit{median
    QMC rules}~\cite{GE22}, that allows for unlimited
    size of integration lattices (previously capped at about
    $7\!\cdot\!10^{10}$), does not sacrifice quality on average,
    and automatically helps with a phenomenon we have named
    \textit{unlucky lattices} that is spoiling the practical
    convergence properties of QMC integration (\ref{sec:medianqmc});
\item
    a new method of automatically introducing the minimal
    set of extra regulators that are required by the
    expansion-by-regions procedure (\ref{sec:extraregulators});
\item
    support for arbitrary arithmetic expressions in the coefficients
    of amplitudes, sparse coefficient matrices, and amplitude
    names (\ref{sec:coefficientsyntax}).
\end{itemize}

\section{The new evaluator \textsc{Disteval}}
\label{sec:disteval}

\begin{figure}
    \centering
    a)~~\raisebox{0.5ex}{\scalebox{0.75}{\begin{tikzpicture}
	\begin{pgfonlayer}{nodelayer}
		\node [style=none] (0) at (-2, 0) {};
		\node [style=dot] (1) at (-1.5, 0) {};
		\node [style=dot] (2) at (-0.75, 0.75) {};
		\node [style=dot] (3) at (-0.75, -0.75) {};
		\node [style=dot] (4) at (0.75, 0.75) {};
		\node [style=dot] (5) at (0.75, -0.75) {};
		\node [style=dot] (6) at (1.5, 0) {};
		\node [style=none] (7) at (2, 0) {};
		\node [style=none] (8) at (-0.25, 0) {$m_W$};
		\node [style=none] (9) at (-1.75, 0.25) {$m_Z$};
		\node [style=none] (10) at (1.75, 0.25) {$m_Z$};
	\end{pgfonlayer}
	\begin{pgfonlayer}{edgelayer}
		\draw [style=incoming massive edge] (0.center) to (1);
		\draw [style=outgoing massive edge] (6) to (7.center);
		\draw [style=massive3 edge] (2) to (3);
		\draw [style=edge] (2) to (1);
		\draw [style=edge] (4) to (2);
		\draw [style=edge] (6) to (4);
		\draw [style=edge] (5) to (6);
		\draw [style=edge] (3) to (5);
		\draw [style=edge] (1) to (3);
		\draw [style=edge] (5) to (4);
	\end{pgfonlayer}
\end{tikzpicture}}} \hspace{1cm}
    b)~\raisebox{0.5ex}{\scalebox{0.75}{\begin{tikzpicture}
	\begin{pgfonlayer}{nodelayer}
		\node [style=none] (0) at (-1.25, -0.75) {};
		\node [style=none] (1) at (-0.5, -1) {};
		\node [style=none] (2) at (-1.25, 0.75) {};
		\node [style=none] (3) at (-0.5, 1) {};
		\node [style=none] (4) at (2, 0) {};
		\node [style=dot] (5) at (-0.75, -0.5) {};
		\node [style=dot] (7) at (0, 0.75) {};
		\node [style=dot] (8) at (0.75, 0.5) {};
		\node [style=dot] (9) at (1.5, 0) {};
		\node [style=dot] (10) at (0.75, -0.5) {};
		\node [style=dot] (11) at (0, -0.75) {};
		\node [style=dot] (12) at (-0.75, 0.5) {};
		\node [style=none] (16) at (-0.25, 0.25) {$m$};
		\node [style=none] (17) at (1, 0) {$m$};
		\node [style=none] (18) at (1.5, 1) {$d=6-2\varepsilon$};
		\node [style=none] (19) at (2, -0.25) {$m'$};
	\end{pgfonlayer}
	\begin{pgfonlayer}{edgelayer}
		\draw [style=outgoing edge] (11) to (1.center);
		\draw [style=outgoing edge] (5) to (0.center);
		\draw [style=outgoing massive edge] (7) to (3.center);
		\draw [style=outgoing massive3 edge] (9) to (4.center);
		\draw [style=edge] (5) to (12);
		\draw [style=massive edge] (12) to (7);
		\draw [style=edge] (7) to (8);
		\draw [style=massive edge, style=dot1] (8) to (9);
		\draw [style=massive edge] (9) to (10);
		\draw [style=massive edge, style=dot1] (10) to (8);
		\draw [style=edge] (10) to (11);
		\draw [style=edge] (11) to (5);
		\draw [style=outgoing massive edge] (12) to (2.center);
	\end{pgfonlayer}
\end{tikzpicture}}} \hspace{1cm}
    c)~\raisebox{0.5ex}{\scalebox{0.75}{\begin{tikzpicture}
	\begin{pgfonlayer}{nodelayer}
		\node [style=dot] (0) at (-0.825, 0.5) {};
		\node [style=dot] (1) at (-0.825, -0.475) {};
		\node [style=none] (2) at (-1.25, 0.5) {};
		\node [style=none] (3) at (-1.25, -0.475) {};
		\node [style=dot] (4) at (0.5, 0.5) {};
		\node [style=dot] (5) at (0.5, -0.475) {};
		\node [style=dot] (6) at (-0.225, -0.475) {};
		\node [style=none] (7) at (0.925, 0.5) {};
		\node [style=none] (8) at (0.925, -0.475) {};
		\node [style=none] (9) at (-0.5, 0.25) {$m$};
		\node [style=none] (10) at (0.925, -0.225) {$m'$};
	\end{pgfonlayer}
	\begin{pgfonlayer}{edgelayer}
		\draw [style=incoming edge] (2.center) to (0);
		\draw [style=incoming edge] (3.center) to (1);
		\draw [style=massive edge] (0) to (1);
		\draw [style=massive edge] (0) to (4);
		\draw [style=massive edge] (4) to (5);
		\draw [style=massive edge] (1) to (6);
		\draw [style=massive edge] (6) to (5);
		\draw [style=edge] (6) to (4);
		\draw [style=outgoing edge] (4) to (7.center);
		\draw [style=outgoing massive3 edge] (5) to (8.center);
	\end{pgfonlayer}
\end{tikzpicture}}}
    \caption{Diagrams used in subsequent figures. Diagrams b and c
    correspond to \texttt{hexatriange} and \texttt{elliptic2L\_physical}
    examples distributed with py\textsc{SecDec}.}
    \label{fig:diagrams}
\end{figure}
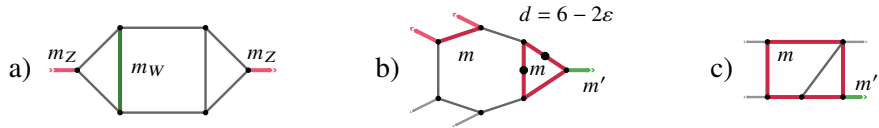

\begin{figure}
    \centering
    \includegraphics[width=0.75\textwidth]{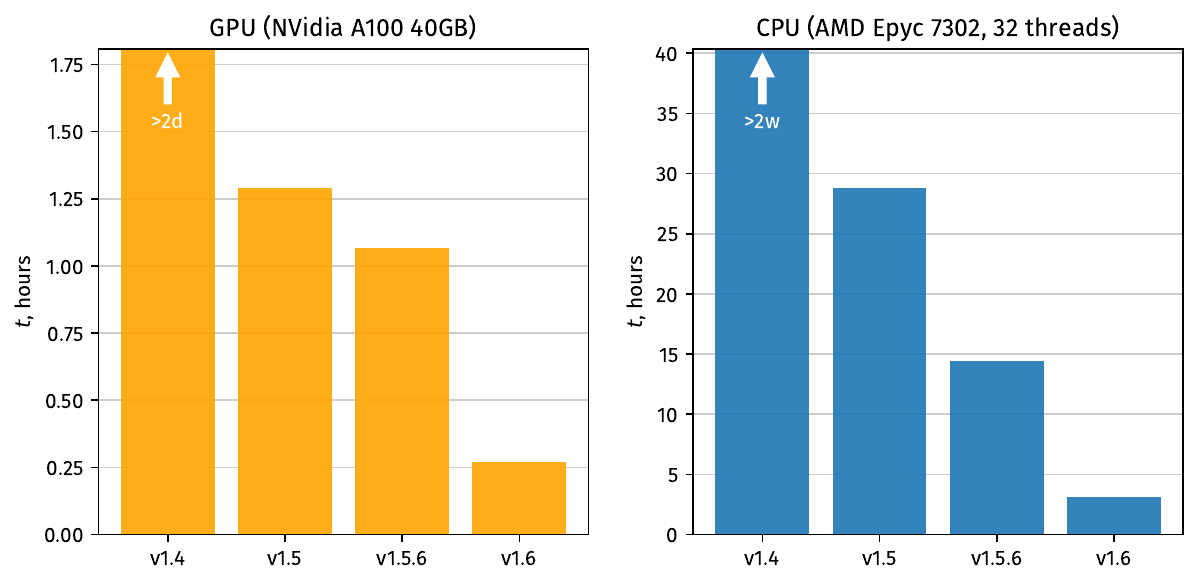}
    \caption{The time it takes different releases of py\textsc{SecDec}
    to integrate \subfigref{fig:diagrams}{a} to 7 digits of precision.}
    \label{fig:perfbyversion}
\end{figure}

\begin{figure}
    \centering
    \includegraphics[width=0.75\textwidth]{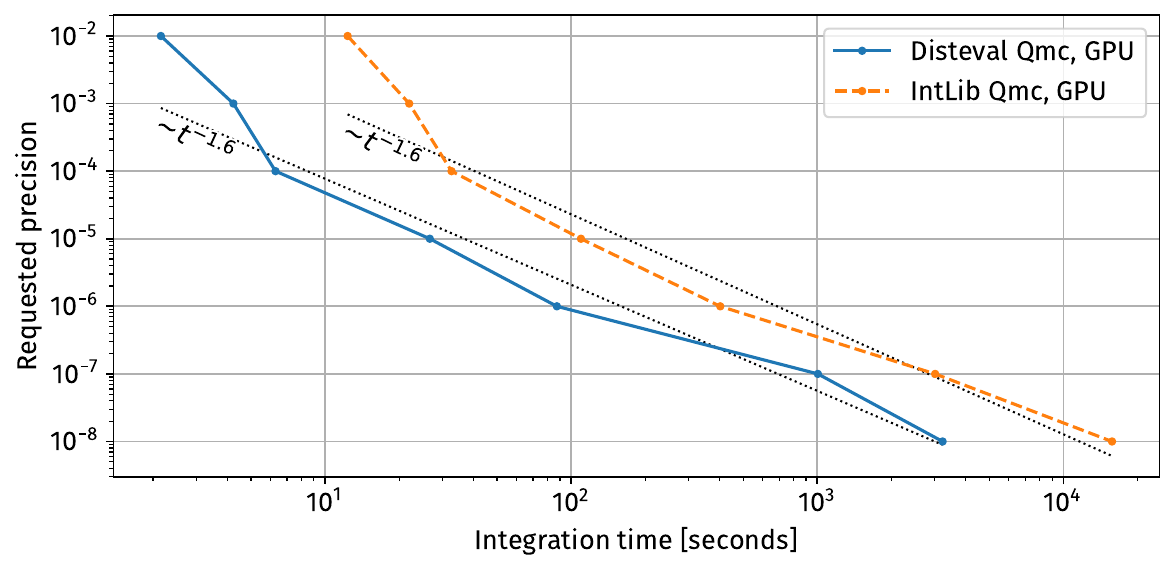}
    \caption{Runtime versus requested precision when integrating
    \subfigref{fig:diagrams}{b} with the old and the new integrators
    of py\textsc{SecDec}~1.6 on NVidia~A100.}
    \label{fig:hextimings}
\end{figure}

\noindent
The first feature of version 1.6 is a major increase in performance.
Over the last few releases py\textsc{SecDec} has been consistently
gaining performance, as can be seen in \ref{fig:perfbyversion}. The
sources of these gains are:
\begin{itemize}
    \item In version 1.5: adaptive sum sampling and automatic contour deformation
        adjustment~\cite{Heinrich:2021dbf}.
    \item In version 1.5.6: microoptimizations in the integrand code
        decreasing the overhead of complex numbers.
    \item In version 1.6: a new Randomized Quasi Monte
        Carlo integrator ``\textsc{Disteval}'', that
        implements the same algorithm as the old integrator
        (``\textsc{IntLib}'') based on the \textsc{Qmc}
        library~\cite{Borowka:2018goh}, but generates code
        that is 3x--9x faster both on CPUs and GPUs. This
        speedup is illustrated on \ref{fig:hextimings} and in
        \ref{tab:hextimings}, and is
        consistent with our benchmarks on many other integrals.
        The speedup comes from many optimizations; the most
        important ones are:
    \begin{itemize}
        \item For CPU and GPU: the code of the integrands is
            generated together with the integrator code, and
            is fully embedded into the integration loop. This
            gives us---and the compiler---many possibilities for
            optimization, such as taking out common expressions
            out of the loop, interleaving integer and floating
            point calculations for better CPU pipeline utilization,
            allocating variables to registers, moving error handling out
            of the hot path, etc.
        \item For CPU: better processor utilization via SIMD
            instructions; in particular the integrands are made
            to operate on four 64-bit values at
            the same time, which translates into higher saturation
            of the execution units on modern CPUs, especially if
            the user has enabled the usage of the AVX2 and FMA
            instruction sets.\footnote{The is
            done by setting \texttt{CFLAGS} to \texttt{"-mavx2
            -mfma"} during compilation. Note however that some
            older CPUs do not support AVX2, which is why instead
            of setting these flags by default we \textit{strongly
            recommend} their usage.}
        \item For GPU: we have largely eliminated stalls due to
            synchronization with the CPU by performing summation of the samples directly
            on the GPU, and by performing most operations asynchronously.
    \end{itemize}
\end{itemize}

\begin{table}

\centering
    \scalebox{0.8}{
\begin{tabular}{llrrrrrr}
\toprule 
\multicolumn{2}{c}{\textsubscript{Integrator}\textbackslash\textsuperscript{Accuracy}} & $10^{-3}$ & $10^{-4}$ & $10^{-5}$ & $10^{-6}$ & $10^{-7}$ & $10^{-8}$\tabularnewline
\midrule
\midrule 
GPU & \textsc{Disteval} & 4.2 s & 6.3 s & 27 s & 1.5 m & 17 m & 54 m\tabularnewline
 & \textsc{IntLib} & 22.0 s & 22.0 s & 110 s & 6.7 m & 50 m & 263 m\tabularnewline
 & Speedup & 5.2 & 5.2 & 4.1 & 5.6 & 3.0 & 4.9\tabularnewline
\midrule 
CPU & \textsc{Disteval} & 5.1 s & 14 s & 1.6 m & 8.3 m & 57 m & 4.7 h\tabularnewline
 & \textsc{IntLib} & 20.8 s & 86 s & 14.2 m & 62.2 m & 480 m & 43.1 h\tabularnewline
 & Speedup & 4.1 & 6.1 & 8.7 & 7.5 & 8.4 & 9.2\tabularnewline
\bottomrule
\end{tabular}\vspace{-3mm}
\par
    }
    \caption{Integration timings depending on requested precision
    on a GPU and CPU corresponding to \ref{fig:hextimings}.}
    \label{tab:hextimings}
\end{table}

\noindent
Additionally \textsc{Disteval} is able to distribute the work
across different computers (hence its name): any computer that
can be reached via \texttt{ssh} can be added to the list of
\textsc{Disteval} workers.

\section{Quasi Monte Carlo lattices via median QMC rules}
\label{sec:medianqmc}

\noindent
For a long time the recommended way of integration with
py\textsc{SecDec} has been Randomized Quasi Monte Carlo integration
method~\cite{DKS:2013} implemented via the \textsc{Qmc}
library. Central to this method is the
construction of the set of points on which to evaluate the
integrands.
For this \textsc{Qmc} uses rank-1 lattices, i.e. sets of
points given by
\begin{equation}
    {\vec x}_i = \left\lfloor \frac{i \, {\vec g}_n}{n} \right\rfloor,
    \qquad
    i = 1 \dots n,
\end{equation}
where ${\vec g}_n$ are the \textit{generating vectors} of the
lattice. These are constructed via the so-called
\textit{component-by-component} (CBC) method~\cite{Nuyens06}.
This construction depends on the function space
the integrands are supposed to belong, and \textsc{Qmc} has been
assuming a Korobov space with smoothness $\alpha = 2$ and product
weights.

This construction however does not guarantee optimal convergence
for any given integral, only for a class of integrals, and only
asymptotically. In practice if one uses a
sequence of lattices of increasing size constructed via the CBC
generating vectors, the integration error does not drop down
monotonically, but instead fluctuates, sometimes very significantly
so.

\begin{figure}
    \centering
    \includegraphics[width=0.75\textwidth]{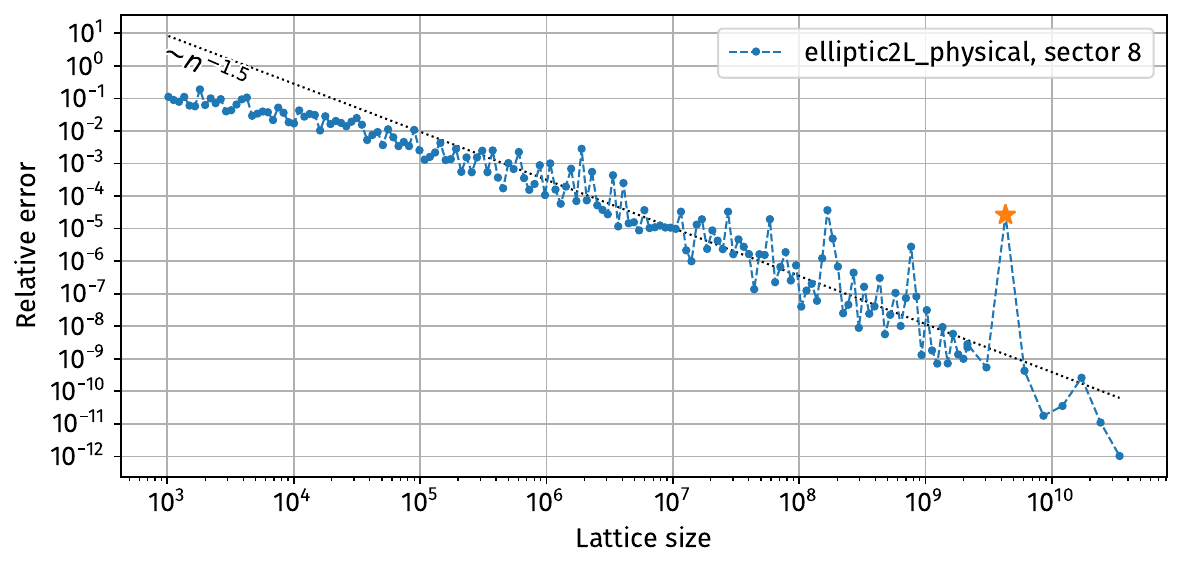}
    \caption{Relative integration error for sector~8 from the
    \texttt{elliptic2L\_physical} example (\subfigref{fig:diagrams}{c})
    achieved by lattices of different size constructed via the
    CBC construction. A particularly unlucky lattice is marked
    with a star.}
    \label{fig:pbl-cbc-sec8}
\end{figure}

Take a look at \ref{fig:pbl-cbc-sec8} for an example: although the
integration error generally scales as $1/n^{1.5}$, some lattices
show much worse errors. We call these \textit{unlucky
lattices}. The result of integration on one such unlucky lattice
with $n\approx4.2\!\cdot\!10^{9}$ is marked with a star---this one
is off the general trend by at least 4~orders of magnitude! If this
particular lattice is chosen during integration, py\textsc{SecDec}
will assume that this integral converges badly, and many
more samples are needed to achieve the requested precision, which
will decrease the overall performance significantly.

\begin{figure}
    \centering
    \includegraphics[width=0.75\textwidth]{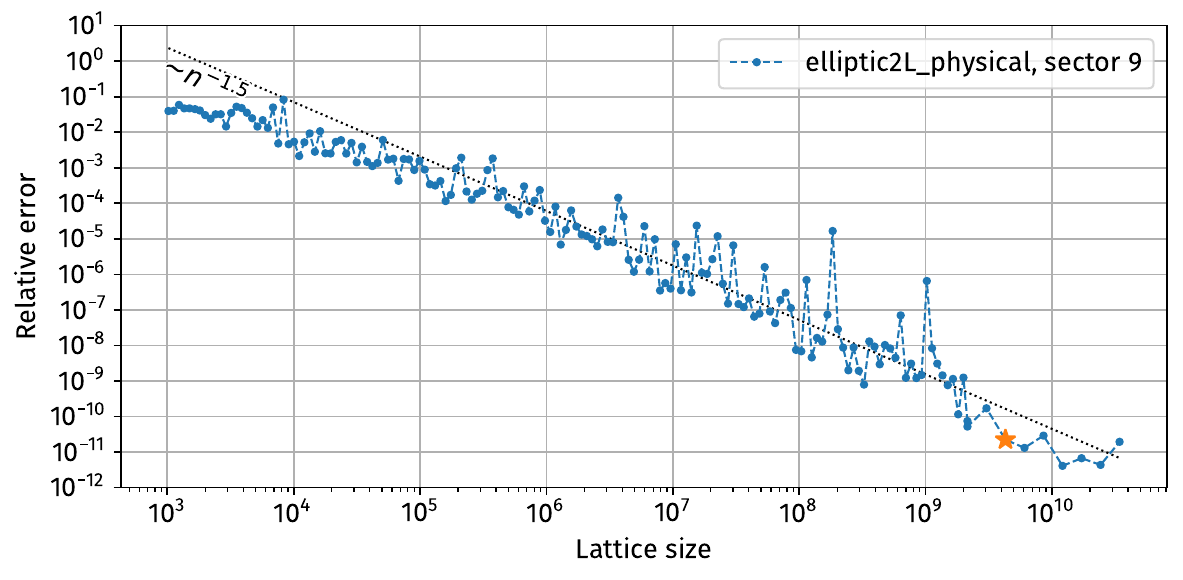}
    \caption{Relative integration error for sector~9 from the
    \texttt{elliptic2L\_physical} example (\subfigref{fig:diagrams}{c})
    achieved by lattices of different size constructed via
    the CBC construction. A star marks the same lattice as in
    \ref{fig:pbl-cbc-sec8}.}
    \label{fig:pbl-cbc-sec9}
\end{figure}

It is important to note that the lattice at $n\approx4.2\!\cdot\!10^{9}$
is not bad per se: rather it is unlucky for a given integrand,
but may be completely normal for others. For example,
\ref{fig:pbl-cbc-sec9} demonstrates the integration error for a
different sector of the same integral as \ref{fig:pbl-cbc-sec8},
and the previously unlucky lattice performs perfectly well.
The source of the problem in our view is that the lattices
are precomputed for a class of integrands (via CBC), rather than
being personalized for each one.

Another problem with the CBC construction is that
it is computationally expensive and \textsc{Qmc}
authors have only been able to construct lattices up to
$n\approx7\!\cdot\!10^{10}$. This size might seem big,
but a server-grade GPU such as NVidia A100 can evaluate
some integrands $10^{10}$ times per minute, and in many examples
reaching integration precisions of e.g. $10^{-8}$
requires more samples than this.

\begin{figure}
    \centering
    \includegraphics[width=0.75\textwidth]{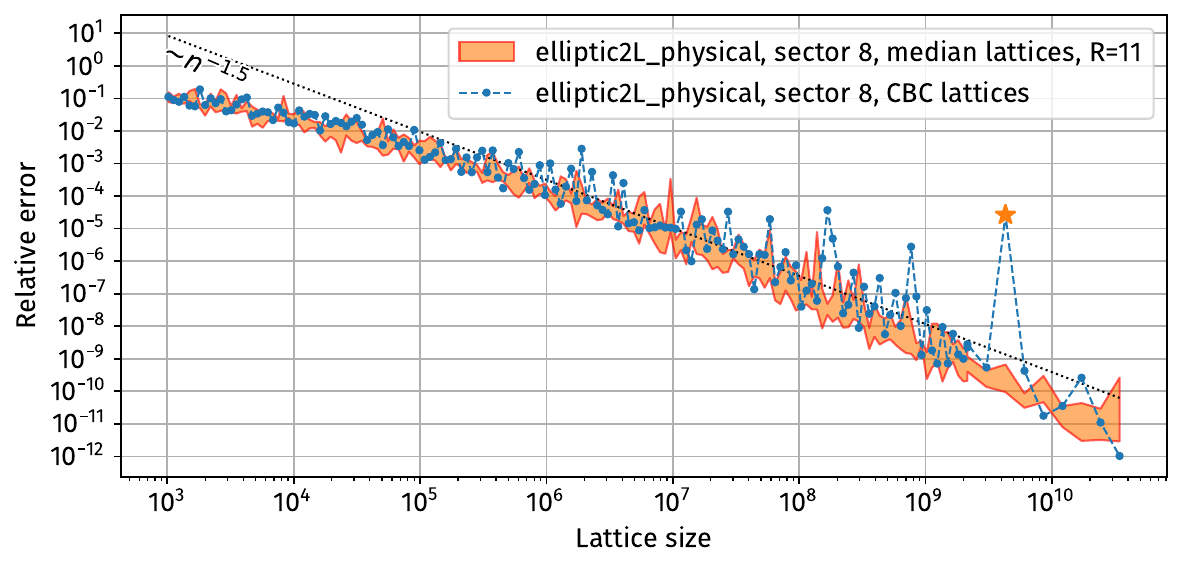}
    \caption{Same as \ref{fig:pbl-cbc-sec8}, but using lattices
    derived via median QMC rules. Because such lattices are
    derived probabilistically (i.e. every time anew), we have
    made 15 measurements for each point, and the bands on this
    plot indicate the range from the worst to the best results.}
    \label{fig:pbl-med-sec8}
\end{figure}

Fortunately, in~\cite{GE22,GSM22} a lattice construction was found
that does not require specific assumptions about the function
space the integrands belong to, and that works for arbitrary
lattice sizes without excessive computations. This construction
works by selecting $R$ random generating vectors, evaluating
the integral on each of the corresponding lattices, and then
accepting the \textit{median} result. This is the \textit{median
QMC rules} construction. It is proven to result in a lattice that
is arbitrary close to an optimal one with a probability that
tends to 1 exponentially with increasing $R$, irrespective of
the smoothness class of the integrand. Our experience confirms
this: take a look at \ref{fig:pbl-med-sec8} where the same
integrand as in \ref{fig:pbl-cbc-sec8} is integrated on lattices
constructed via the median lattice construction; as can be
seen the convergence is on average as good as the one with CBC
lattices; moreover the extremely unlucky lattices are avoided.

The median QMC rules construction is now available in
py\textsc{SecDec}, both with the new and the old integrators.
Since it is rather new, it is not yet the default, but it can
be activated by specifying the number of lattice candidates~$R$
via the \texttt{lattice\_candidates} parameter.

\section{Minimal extra regulator construction for expansion by regions}
\label{sec:extraregulators}

\noindent
Since version 1.5 py\textsc{SecDec} has the ability to
perform asymptotic expansions~\cite{Heinrich:2021dbf}
(i.e. expansions by regions~\cite{Smirnov:1991jn,Beneke:1997zp,Pak:2010pt,Jantzen:2011nz})
via the \texttt{loop\_regions()} function.
It is known that this procedure can introduce spurious
singularities, not regulated by the original dimensional regulator.
These show up as $1/0$ division errors, and are commonly regulated
by either shifting the powers of the propagators from their
original values by some infinitesimal~$\nu_i$, or by introducing
an additional factor of the form $\prod_i x_i^{\nu_i}$ into the
Feynman parameterization. In both cases $\nu_i$ become extra
regulators---or rather, they are set to $\nu/2$, $\nu/3$, $\nu/5$,
etc, and $\nu$ becomes a single extra regulator.

For example, if one wishes to expand a box integral, 
\begin{equation}
    I(m^2,s,t) \equiv
    \raisebox{0.5ex}{\scalebox{0.75}{\begin{tikzpicture}
	\begin{pgfonlayer}{nodelayer}
		\node [style=dot] (0) at (-0.75, 0.75) {};
		\node [style=dot] (1) at (0.75, 0.75) {};
		\node [style=dot] (2) at (-0.75, -0.75) {};
		\node [style=dot] (3) at (0.75, -0.75) {};
		\node [style=none] (4) at (-1, 1) {};
		\node [style=none] (5) at (1, 1) {};
		\node [style=none] (6) at (-1, -1) {};
		\node [style=none] (7) at (1, -1) {};
		\node [style=none] (8) at (0, 0) {$m$};
		\node [style=none] (9) at (-1.25, 1) {$p_1$};
		\node [style=none] (10) at (-1.25, -1) {$p_2$};
		\node [style=none] (11) at (1.25, -1) {$p_3$};
		\node [style=none] (12) at (1.25, 1) {$p_4$};
		\node [style=none] (13) at (0, 1) {$1$};
		\node [style=none] (14) at (-1, 0) {$2$};
		\node [style=none] (15) at (0, -1) {$3$};
		\node [style=none] (16) at (1, 0) {$4$};
	\end{pgfonlayer}
	\begin{pgfonlayer}{edgelayer}
		\draw [style=outgoing edge] (0) to (4.center);
		\draw [style=outgoing edge] (1) to (5.center);
		\draw [style=outgoing edge] (3) to (7.center);
		\draw [style=outgoing edge] (2) to (6.center);
		\draw [style=massive edge] (0) to (1);
		\draw [style=massive edge] (1) to (3);
		\draw [style=massive edge] (3) to (2);
		\draw [style=massive edge] (2) to (0);
	\end{pgfonlayer}
\end{tikzpicture}
}} =
    \int \mathrm{d} x_1 \cdots \mathrm{d} x_4 \, U^{\alpha} (\vec x)
    F^\beta (\vec x, m^{2}, s, t) \delta(1-x_1-\cdots-x_4),
\end{equation}
asymptotically in small $m^2/s$ ratio, extra regulators
$\nu_1,\dots,\nu_4$ might be introduced like so:
\begin{equation}
    \mathrm{ebr}\!\left[\,I\,\right]=
    \lim_{\nu_{1,2,3,4}\,\to\,0}\,
    \mathrm{ebr}\!\left[\int 
    \mathrm{d} x_1 \cdots \mathrm{d} x_4\,U^{\alpha}\,F^{\beta}\, \delta (1-x_1-\cdots-x_4)
    \cdot{x_1^{\nu_1}x_2^{\nu_2}x_3^{\nu_3}x_4^{\nu_4}}\right],
    \label{eq:box-ebr}
\end{equation}
and the values of $\nu_i$ could be set to e.g. $\{\nu, \nu/2, \nu/3, \nu/5\}$.

The introduction of extra regulators has a negative consequence
in that the integral becomes more complicated, and that many
symmetries it might have had otherwise are spoiled, because
each propagator now has a unique exponent. For this
reason, when calculating amplitudes consisting of many integral,
two questions are of practical interest:
\begin{enumerate}
    \item Can we detect if an integral can be expanded as is,
        or if it needs extra regulators?
    \item If it does need them, can we detect the minimal set
        of propagators that must be spoiled, as to avoid spoiling
        the rest?
\end{enumerate}
The answer to both is ``yes'', and~\cite{Heinrich:2021dbf}
provides a geometric construction of detecting the need for extra
regulators. Specifically, a set of vectors ${\vec n}_j$ can be
derived for any given integral, such that for extra regulators
$\vec \nu \equiv \{\nu_i\}$ the integral is sufficiently regulated
if ${\vec n}_j\cdot{\vec \nu} \ne 0 \ (\forall j)$. Starting
with py\textsc{SecDec} 1.6 we provide an implementation of this
construction via functions \texttt{extra\_regulator\_constraints()}
and \texttt{suggested\_extra\_regulator\_exponent()}.

To continue the example of \ref{eq:box-ebr}, one can use
\texttt{extra\_regulator\_constraints()} to obtain the set of
necessary conditions on $\nu_i$. In this case those turn out to be
\begin{equation}
    \nu_{2}-\nu_{4}\ne0 \quad\text{and}\quad\nu_{1}-\nu_{3}\ne0.
\end{equation}
Then, \texttt{suggested\_extra\_regulator\_exponent()} will give a
suggested solution to these inequalities that maximizes the number of
$\nu_i$ set to zero; in this case it is
\begin{equation}
    \nu_i=\{ 0,\,0,\,\nu,\,-\nu\}.
\end{equation}
The overall result here is that instead of blindly introducing an
extra factor like $x_1^{\nu/1}x_2^{\nu/2}x_3^{\nu/3}x_4^{\nu/5}$,
we only need to insert $x_3^{\nu}x_4^{-\nu}$, and the integral
will be well regulated after expansion by regions.

This procedure can now be performed automatically by py\textsc{SecDec}:
if the user specifies the \texttt{extra\_regulator\_name} argument
to \texttt{loop\_regions()} without giving a corresponding
\texttt{extra\_regulator\_exponent}, an extra regulator will be
automatically introduced in a minimal way, or even skipped if
the integral does not require it.

\section{More general syntax for amplitude coefficients}
\label{sec:coefficientsyntax}

\noindent
Since version 1.5 py\textsc{SecDec} comes with an interface
to evaluate weighted sums of integrals (i.e amplitudes) in an
optimized way: the \texttt{sum\_package()} function. Previously
this interface required the coefficients to be passed in as
products of polynomials in the dimensional regulator, however this
format is inconvenient in practice, as the coefficients often
come from integration-by-parts reduction as large expressions too
complicated for further transformations.

In py\textsc{SecDec} 1.6 the format of the coefficients is
relaxed, and they can be passed in as strings containing arbitrary
arithmetic expressions. These expressions are then parsed
and evaluated via \textsc{GiNaC}~\cite{Bauer:2000cp} during
integration, and with \textsc{Disteval} this is even done in
parallel, allowing for large number of coefficients (i.e. large
amplitudes) to be handled efficiently. In the same vein we we
avoid loss of numerical precision during intermediate calculation
by performing the evaluation in infinite precision arithmetics.

Additionally, \texttt{sum\_package()} now accepts coefficient
matrices as dictionaries of the form\footnote{The details of the usage are available in the py\textsc{SecDec}
documentation. The new syntax is also illustrated in the
\texttt{easy\_sum} and \texttt{muon\_production} examples
in the source repository.}
\begin{equation*}
    \text{\texttt{\{"}\textit{amplitude name}\texttt{":
    \{}\textit{integral index}\texttt{: "}\textit{coeffcient}\texttt{",
    }\textit{...}\texttt{\}, }\textit{...}\texttt{\}},}
\end{equation*}
making it possible to give amplitudes names, and
to efficiently specify sparse coefficient matrices, since
zero coefficients can be skipped in this notation.


\section{Conclusions}

\noindent
py\textsc{SecDec} release 1.6 comes with multiple features
targeted at faster and easier evaluation of amplitudes and
single integrals. The new integrator \textsc{Disteval} brings a
significant performance increase on both the CPU and the GPU.
A new QMC lattice construction allows for higher sample counts
(meaning higher precision), and comes with builtin unlucky lattice
mitigation---also needed at high integration precisions. Support
for arbitrary arithmetic expressions in amplitude coefficients
enables users to easily specify the coefficients of arbitrary
size without preprocessing. An automatic and minimal construction
of extra regulators for expansion-by-regions gives a streamlined
path to asymptotic expansion of large number of integrals without manual
interventions.

We hope that these features will prove useful to the audience
at large, and will enable calculations in high energy physics
previously deemed too complicated. We also hope to continue
improving py\textsc{SecDec} for the benefit of the community.

\section*{Acknowledgements}

\noindent
This research was supported in part by Deutsche Forschungsgemeinschaft
(DFG, German Research Foundation) under grant 396021762 - TRR~257.

\bibliographystyle{JHEP}
\bibliography{main}
\end{document}